Short Paper

# A Review of Game-based Mobile E-Learning Applications

Carlo H. Godoy Jr.
College of Industrial Technology, Technological University of the Philippines
carlo.godoyjr@tup.edu.ph

**Abstract**

*Purpose* – This study aims to review and get information on the different mobile game applications and the possibility of being a supplementary tool for learning to enhance and empower the e-learning aspect of a Technical Education and Skills Development Authority (TESDA). It is reviewed to help in the development of the TESDA's E-Learning department. The study aims to facilitate the re-organization of the TESDA's E-learning program by using a method called Game-Based Learning. To accomplish this target, different applications will be reviewed to give an idea on how game-based learning is being used as a supplementary learning tool for different subject areas. This paper will focus on TESDA's language Skills institute and Technical-Vocational program since there are a lot of reviews for main subject areas but only few in this field.

*Method* – The database to be used in the methodology is Google Scholar, inclusion and exclusion criteria used is that article should be within the span of five years and should be game based learning mobile applications.

*Results* – Will be selecting 4 applications as a representation since it is an informational review. No need to indicate the other applications as this is just a basis for future studies.

*Conclusion* – Mobile games have such a powerful instructional potential, by which teaching can be revolutionized, which output is acquiring the praise and judgment of academic professionals and educators.

*Recommendations* – After this review, future studies can be made to support this study like reviewing from other databases like Web of Science and Scopus to support the foundation made on this review.

*Research Implications* – The review will open further studies to help TESDA administrator in using mobile games as a supplementary learning tool for their e-learning.

*Practical Implications* – TESDA would be able to get an idea on how Technical-Vocational and Language Skills can be taught alternatively in the new normal.

*Social Implications* – Since social distancing is being implemented, students can still enjoy and communicate by using mobile game as an avenue for socialization.

## INTRODUCTION

E-learning through mobile game application is basically associated with game-based learning. Educational games have been described as applications which use video-related mobile game features to build engaging and immersive learning experiences with specific goals. These games generate challenges, promote different levels of communication and provide fun multimedia and immediate feedback (Denden, Essalmi & Tlili, 2017). Letting the players take advantage of the gameplay to achieve certain goals would make the players be more motivated to play the game as the rule will ultimately make them feel better once the goal is achieved. The perspective that is needed for games to achieve the full potential is as follows: affective, cognitive, socio-cultural and motivational (Plass, Homer & Kinzer 2015). All these views need to be taken into account, with special emphasis depending on the purpose and design of the learning method as well as the game itself.

In terms of assessing the result, using the experience points of the players in a certain mobile application game as a measure of good learning would allow them to understand the reason of getting a low score and use it to have a better game play and learning experience after repeating the game. Adaption, on the other hand, uses computer-generated models in the game as a replica to the situations from the real world for the game to be adapted as a good learning game since it uses scenario-based approach. Garris (2002) (as cited by Ke, Xie & Xie 2016) stated that good learning games are anticipated to involve gamers in a learning method that is problem-based. This is where the players will try to decide on integrated questions, experiment on the alternatives offered or strategies, the feedback of the system is being interpreted, reflect on and adaptation of techniques for the development of fresh insights or skills.

This study aims to get proper information about on how mobile game applications are being used for e-learning. The main goal of this review is to present different mobile game applications being used as supplementary learning tools in selected areas of education, in this study. of using mobile application for e-learning is to prove the feasibility of applying a mobile application as an aid to learn difficult subjects like pre-calculus, chemistry, physics, and many others. To accomplish this target, the proponents need to collect essential basic information, research publications and projects, patents, and data from Google Scholar.

This study will provide information on how game-based learning thru mobile app can help in the aspect of establishing an e-learning by the aid of a supplementary tool. This method will serve as an advantage to help students have a fun yet productive learning experience.

## THEORIES AND METHODS USED IN EDUCATIONAL MOBILE GAMES

To promote learning and development, it is certainly a big help to use games as an instructional tool for learning context within the classroom. In the classroom, professional development should communicate the relationship between social learning theory and game-based learning (Polin, 2018). New virtual environments should emerge in the mainstream culture, which may be useful for both learning and entertainment. According to the Constructivist philosophy, thinking takes place by activating one's thoughts and helping to focus on them. This approach helps learners to understand how fresh ideas, actions taken and interactions make sense of their own mental models (Reeve, 2012). Some learners have generally revealed that academic learning through games is linked to a wide range of causes and effects of behavior, affective, perceptual, cognitive and motivational, but it also shows a wide range of research questions, methods, game types, theories and designs (Stiller & Schworm, 2019). This perspective is derived from the behaviorism and constructivist theory.

Educational game is one of game-based learning methods to be applied. Educational games have been described as apps that use video-related mobile game features to create interactive and immersive learning experiences with specific goals. These games create challenges, foster different levels of interaction, and provide pleasant multimedia and immediate feedback (Denden, Essalmi & Tlili 2017). In many areas such as language learning and mathematics, studies have shown the usefulness of educational games. Several scientists have discovered that the narrative meaning is an important aspect of successful educational game design. According to Dickey (as cited by Derwich & Essalmi 2017), it offers a conceptual structure for students to solve their problems as the narrative plot in some games creates an environment in which players will be able to recognize and build trends that are known to combine what is casual with what is highly speculative but acceptable in the deep meaning of the story. Another important component of efficient educational game design is objectives as well as rules of play. Although incorporated in a story frame, goals and rules are not subject to context; they are similarly important components of context.

A way to create a game that focuses primarily on learning a specific topic is to introduce the concept of a smart game. It is considered complicated to design a Smart Game Based Learning System (SGBLS) (Mejbri, Khemaja, & Raies 2017). It needs to interfere with different actors with specific skills and knowledge. Unfortunately, novice game developers who do not have the necessary skills inspired by instructional and video games systems cannot create SGBLS effectively. The overlap of various features of pervasive games-based smart learning systems (PGBSLSs), including the pervasive aspect, ludic aspect and academic aspect, contributes to the complexity of the use of design time and runtime standards (Mejbri, Khemaja, & Raies 2017). These methods have been regarded not only in the gaming industry but also in the education industry as an innovation technology. Innovation technology has been seen as a promising alternative to learning and teaching in recent years (Quadir et al., 2017). To promote more effective learning and teaching, such as Web 2.0 (Steel & Levy 2013), computer-mediated communication (Sun & Yang, 2015), and game-based learning (Escudeiro & Carvalho 2013), these innovative techniques have been developed.

Mobile game applications also have the element of a video game since it is using visuals to entice the player to play the game. One good example is an Augmented Reality Mobile Game which uses visuals to get the attention of the player (Godoy Jr, 2020). Augmented Reality (AR) apps have received increasing attention over the previous two decades. In the 1990s, AR was first used for apps linked to pilot education as well as for training of Air Force (Akçayır & Akçayır 2016). AR generates fresh world experiences with its data layering over 3D space, suggesting that AR should be embraced over the next 2–3 years to give fresh possibilities for teaching, learning, study, or creative investigation according to the 2011 Horizon Report (Chen et al., 2017). AR uses virtual objects or data that overlap physical objects or environments to create a mixed reality in which virtual objects and actual environments coexist in a meaningful manner to increase learning experiences. Azuma et al. (2001) (as cited by Akçayır & Akçayır, 2016) stated that the mentioned virtual objects is appearing in coexistence as the same space as the objects that is located in the real world. AR is now a common technology commonly used in instructional environments in the education sector (Fernandez, 2017).

AR has also become a major study focus in latest years. One of the most significant factors for the widespread use of AR technology is that it no longer needs costly hardware and advanced machinery such as head mounted screens (Akçayır & Akçayır 2016). Azuma (2004) (as cited by Yilmaz, 2016) stated that Augmented Reality is described as having the following characteristics: integrating actual live environment with computer created environment, offering conversation as well as showing 3D items. All of the mentioned components can really be helpful to develop psychomotor skills of vocational trainees through simulation method. By using simulators, trainees can easily replicate the methodologies of a certain industrial based training which is very needed in the Technical-Vocational programs of TESDA.

# RESEARCH METHODOLOGY

The searching procedure started by selecting the topic to be reviewed (Figure 1). In this case the topic is should be related to applications being used as Game Based Learning application. After identifying the topic, the next step is to go to Google Scholar. Google Scholar will be the sole database to be explored in this study. The reason for using Google Scholar is mainly because according to Zientek et al. (2018), (1) the study can easily be tracked using the Google Scholar Profile; (2) Google Scholar can easily aid in the identification of collections of different publications for a specific research topic; (3) Google Scholar lets a researcher to easily track the research overtime for a publication or researcher; (4) it promotes meta-analytic research; (5) it normally bridges social media and scholarly research.

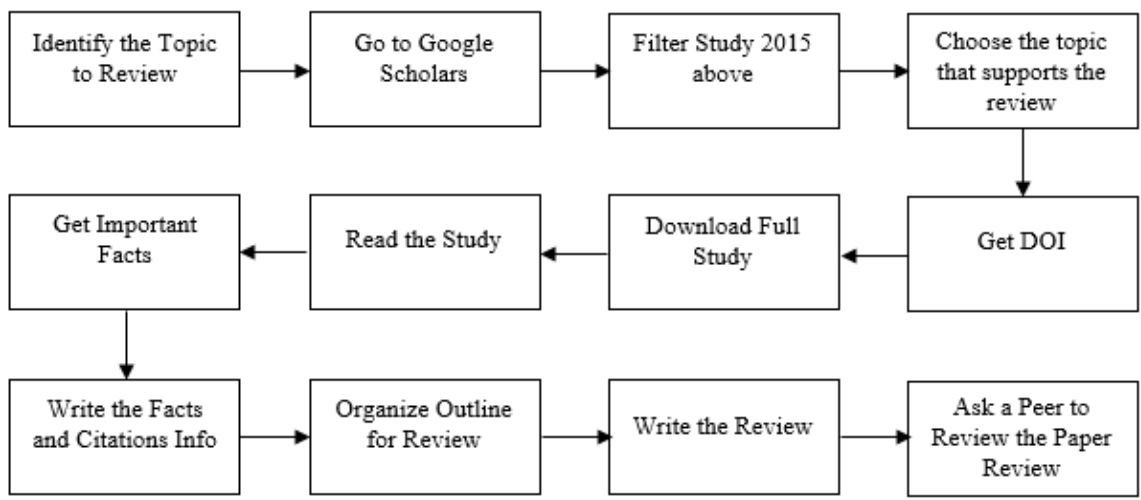

Figure 1. Overview of Research Methodology

The study will be filtered depending on the importance of each study to show how mobile game is being used already as a supplementary tool. Another filter that needs to be added is the year when the publication was published. It is very safe to say that five years interval will still make a certain publication still valid. Lastly, selecting a few applications for Technical-Vocational and Language Skills is also a filter that needs to be added. Once the filtering has been set, it is now time to select the relevant document that will build up the foundation of the review. After knowing the foundation and the outline where the documents will be discussed, the DOI will be used to be able to get the full copy of the documents. Once the full study has been acquired, it is now time to review all documents. During the review process, this is the time to filter which documents are needed to support the selected topic. Take down notes and get all the helpful information for the citation. The review of the mobile game apps for education was written after the information has been gathered and studies have been selected.

Table 1. Exclusion and Inclusion Criteria

| Exclusion Criteria | Inclusion Criteria |
|---|---|
| **Comparison of GBL to normal and blended learning** | Effectiveness of games for Educational Purpose as a supplementary tool |
| **Games that are not for SPED, Language Skills and Technical Vocational Education** | The study should either be for SPED, Language Skills and Technical Vocational Education |
| **Duplicate Studies** | The game should be visually impacting in terms of the subject area. |

While searching on Google Scholar the inclusion and exclusion criteria will be used. In this case, the year when the publication was published should be between 2015 and 2020. So, this means that the study beyond and before the mentioned range should be excluded. The study will be filtered as well depending on the importance of each study to show how mobile game is being used already as a supplementary tool and its effectiveness. This study should show how the tool helps in learning the selected topic not it is being compared with the traditional style of learning. Selecting a few mobile applications for Technical-Vocational and Language Skills which includes SPED Education to represent the review is also considered to avoid bias selection. Lastly, duplicate studies should be excluded, and the game should be visually impacting in terms of learning SPED, Technical-Vocational and Language Skills

## RESULTS AND DISCUSSION OF DIFFERENT MOBILE BASED GAME

Several mobile game applications are used nowadays to supplement teaching and learning process of Language and Technical Vocational subjects. For Technical-Vocational google scholar returned a total of 1660. For Language skills, google scholar returned 143, 000. After applying the inclusion and exclusion matrix illustrated in Table 1, below are selected few as a representation.

1. ChronoOps

    An Augmented Reality based game called ChronoOps has been used to scientifically test the behaviors of language learners. An scientific study of language students involved in using an AR location-based portable match that introduces situational as well as encouraging respondents to grow beyond the traditional subject roles connected with ' student ' or ' learner ' roles is the focus of the ChronoOps (Thorne & Hellermann, 2018). Researchers on this study used the terms hypercontextualization and situated usage events as a result

of their empirical analyzes to describe the intentional structuring of language learning opportunities that occur during a mobile place-based AR game. Multimodal analysis based on EMCA shows the way participants index in their immediate physical context and makes relevant material resources.

These results endorse AR place-based task design as a way to promote the use of the immediate context and the physical environment as a raw material for improvisational and collaborative achievement of AR tasks by the participants. The study made with chronoops shows that a mobile game application is also effective in language study.

2. Fancy Fruits

Another Augmented Reality based application has been create for special needs education called "Fancy Fruits." It is used to teach children with disability the components of regional vegetables as well as regional fruits. The app contains marker-based AR components that connect with virtual data to the actual scenario. A field survey was carried out to assess the request. The research was attended by eleven kids with mental disabilities. The findings indicate that the respondents has a high level of pleasure. Outcomes from a field study demonstrate the beneficial potential of the app: a great joy of use and a child-friendly design. Since researchers of fancy fruit interviewed children with intellectual disorders (Steinhaeusser et al., 2019).

3. Paint-cAR

In Technical and Vocational Education and Training (TVET) organizations, educators see significant challenges on learning system owing to a broad range of SPED necessity of learners a. A marker-based mobile Augmented Reality app called Paint-cAR has been created in aiding the method of teaching fixing car paint as included in vehicle maintenance vocational training program (Bacca et al., 2015). The application was created using a methodology and principle of Universal Design for Learning (UDL) to aid or assist deeply in the development of portable augmented apps in instructional Collaborative creation purposes. To validate Paint-cAR application in a true situation, a cross-sectional assessment survey was performed.

As for the outcome, the inclusive learning design of the AR program does not only help students with special educational needs, but all students will also take advantage of a successful design. This means AR will help solve some of the one-size-fits-all curricula obstacles and promote expert learning. The incorporation of students, teachers, educational technology experts and software developers into a collaborative development (co-creation) process could achieve successful design.

4. Explorez

   In terms of French language, a learning tool called Explorez has been developed. Explorez enables learning to happen outside the classroom with the objective of offering a contextual and immersive educational experience: one that is important and applicable to the students (Perry, 2015). This application proves that Augmented Reality works as an e-learning instrument for enhanced comprehension of content, learning spatial constructions, language connections, long-term memory retention, enhanced cooperation, and motivation.

## CONCLUSIONS AND RECOMMENDATIONS

Digital games have a powerful teaching potential, in an extent education can be revolutionized through it, which output is obtaining academic professionals as well as educators ' praise and judgement. Teachers are a critical component of the program that encompasses educational games creation and use. Mobile games allow school teachers to move their training outside the classroom and connect their teaching with student learning using meaningful items and settings outside the school (Huizenga et al., 2019). More and more schools, educators and learners have access to various kinds of technology and media in recent years, leading to technology-enhanced learning (TEL) being of paramount significance to educators, technology developers and policy makers.

Among these systems, Augmented Reality (AR) is a technological strategy that offers apps that enable learners to communicate with the actual globe through virtual data, and Game-Based Learning (GBL) is a pedagogical strategy that promotes the use of learning games to sum up all preceding discussions. Combining the two process will result to a new system that will give a big impact in the education industry (Tobar Muñoz, 2017). After this review, future studies can be made to support this study like reviewing from other databases like Web of Science and Scopus to support the foundation made on this review.

## IMPLICATIONS

This research revealed that a mobile game app that has a visually impacting representation like Augmented Reality increases the motivation of TVET, SPED and Language Skills students especially in the dimensions of trust and satisfaction since the real world is integrated with 3D designs. When students are in a real environment (like a workshop) with real objects and are driven by increased knowledge, it tends to be an activity in which trust and satisfaction are increased, thereby increasing motivation. Hence this proves that a mobile game application can be used as an aid in teaching TVET, SPED and Language Skills subjects.

**Author's Biography**

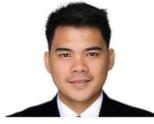 Mr. Carlo H. Godoy Jr is certified Fortinet's Associate Network Security Engineer (NSE). He is also a Support Analyst for SQL at Human Edge Software Philippines. He is a former Escalations Manager at Novartis Pharmaceutical. A Research Scholar and a graduating master's in information technology student at Technological University of the Philippines specializing in studies about Emerging Technologies. He has several research projects on Augmented Reality and Microprocessor Based Systems.

ORCID: https://orcid.org/0000-0002-7701-8036

Web of Science ResearcherID: AAO-2785-2020